\documentclass{revtex4}
\usepackage{natbib}
\usepackage{graphicx}

\begin{document}
	\title{Weyl node assisted conductivity switch in interfacial phase change memory with van der Waals interfaces}
	
	\author{Jinwoong Kim}
	\affiliation{Department of Physics and Astronomy, California State University, Northridge, CA, USA}
	
	\author{Jeongwoo Kim}
	\affiliation{Department of Physics, University of California, Irvine, USA}
	
	\author{Young-Sun Song}
	\affiliation{Department of Physics, Pohang University of Science and Technology, Pohang, Korea}
	
	\author{Ruqian Wu}
	\affiliation{Department of Physics, University of California, Irvine, USA}
	
	\author{Seung-Hoon Jhi}
	\affiliation{Department of Physics, Pohang University of Science and Technology, Pohang, Korea}

	\author{Nicholas Kioussis}
	\email{nick.kioussis@csun.edu}
	\affiliation{Department of Physics and Astronomy, California State University, Northridge, CA, USA}
	
	\date{\today}

\begin{abstract}
The interfacial phase-change memory (iPCM) GeTe/Sb$_2$Te$_3$, promising candidates for the next generation non-volatile random-access memories, exhibits fascinating topological properties. Depending on the atomic-layer-stacking sequence of the GeTe block, the iPCM can be either in the SET (Ge-Te-Ge-Te) or RESET (Te-Ge-Ge-Te) states, where the former exhibits ferroelectric polarization and electrical conductivity two orders of magnitude larger than that of the RESET state. But, its origin remains elusive. Here, we predict the emergence of a Weyl semimetal phase in the SET state induced by the ferroelectric polarization which breaks the crystal inversion symmetry. We show that the giant conductivity enhancement of the SET phase is due to the appearance of gapless Weyl nodes. The Ge-Te- or Sb-Te-terminated surfaces of Weyl semimetal iPCM produce surface states with completely distinctive topology, where the former consists solely of Fermi arcs while the latter consists of both closed Fermi surface and open Fermi arcs.
The iPCM with van der Waals interfaces offers an ideal platform for exploiting the exotic Weyl properties for future memory device applications. 
\end{abstract}

\maketitle

\textbf{Phase change memory.}
Among the chalcogenide phase-change materials, Ge$_2$Sb$_2$Te$_5$ (GST), continues to be one of the most attractive subjects for active experimental and theoretical investigations because of its extensive use in optical memory devices (DVD-RAM), as well as its high potential for non-volatile phase change memory (PCM) devices\cite{GSTa,GSTb,GSTc, GST_TIa, GST_TIb}.
Both applications rely on the large contrast in optical and electric properties between the crystalline and amorphous phases of the phase-change materials which can be switched reversibly and rapidly by sweeping temperature either via laser irradiation (DVD) or the Joule effect (PCM).

\textbf{Interfacial phase change memory.}
Recently, a new type PCM structure consisting of a two-component (GeTe)$_n$/(Sb$_2$Te$_3$)$_m$ superlattice was proposed, referred to as interfacial phase change memory (iPCM)\cite{ipcm}.
It offers reduction in switching energies, improved write-erase cycle lifetimes, faster switching speeds and
surprisingly large room-temperature magnetoresistance even without any magnetic element.
It is noticeable that the two end members of the iPCM exhibit distinct properties; Sb$_2$Te$_3$ is a three-dimensional topological
insulator (TI) with a band inversion\cite{HZhang} at the $\Gamma$-point, while GeTe is a narrowgap (0.55 eV) ferroelectric material with a giant Rashba effect\cite{Picozzi}. The ferroelectricity of GeTe persists down to the nanometer scale\cite{Polkinga,Polkingb}.
Compared to their PCM counterparts, iPCM
remains in the crystalline phase in both the high- (SET) and low-conductivity (RESET) states, and the switch process involves a one-dimensional atomic motion normal to the interface. 
More explicitly, the stacking sequences within the GeTe film in the SET and RESET phases are $\ldots$-Te-Ge-Te-Ge-$\ldots$ and
$\ldots$-Te-Ge-Ge-Te-$\ldots$, respectively. While the former exhibits ferroelectric polarization due to the breaking of the crystal inversion symmetry\cite{ipcm, TominagaSTAM}, the later has no polarization.
The underlying mechanisms of the phase change dynamics and the giant change of resistance between the SET and RESET states still remain elusive\cite{ipcmb}.

One important property of the RESET state of the iPCM is the topologically nontrivial order inherited from the bulk Sb$_2$Te$_3$, although the band inversion strength is somewhat reduced by the intercalating GeTe block\cite{ipcmc}. In the SET state, the GeTe block not only reduces the band inversion strength but also breaks the spatial inversion symmetry.
Consequently, this raises a question whether the SET phase is a Weyl semimetal. It was proposed \cite{murakamia,murakamib} that the absence of the inversion symmetry guarantees the Weyl phase to emerge at the phase boundary between topological and trivial insulator phases.

\textbf{Weyl semimetal.}
Weyl semimetals (WSMs) exhibit a plethora of unusual and intriguing properties.
WSMs are three-dimensional crystals, which break either time-reversal or inversion symmetry, where pairs of energy bands touch at certain $k$ points, referred to as Weyl nodes\cite{Weyl,WeylNN}, and the bands disperse linearly in the vicinities around the Weyl nodes.
The Weyl nodes have chiralities (topologically protected chiral charges) and act as sources or sinks of the Berry curvature flux in the Brillouin zone (BZ)\cite{WeylNN,Savrasov2011,Balents2011,Ran2011}.
WSMs host unusual surface states which form disjoint open Fermi arcs rather than closed Fermi curves, connecting the projections of the Weyl nodes onto the surface BZ where the Fermi arcs on opposite surfaces of the sample are linked through the bulk states\cite{Savrasov2011,Ran2011}.

Despite extensive search for WSMs, the most convincing signature of WSM has been observed only in the covalently-bonded TaAs family\cite{TaAsa, TaAsb}.
To prevent the intervention of surface dangling bonds for the observation of the Fermi arcs, it is typical to find materials whose layers are weakly bonded by the van der Waals forces. The ferroelectric insulator BiTeI with van der Waals layers was predicted to undergo a phase transition
from normal insulator to topological insulator with an intervening Weyl semimetal phase under a hydrostatic pressure\cite{VanderbiltPRB2014}.
However, the pressure window is too narrow to realize steadily the Weyl semimetal phase.
	
\textbf{Objective.}
Here, employing {\it ab initio} electronic structure calculations we predict that the SET phase of the van der Waals type iPCM structure is a Weyl semimetal.
This is due to the ferroelectric polarization in the GeTe building block that breaks the inversion symmetry. We also elucidate that the underlying mechanism of the giant enhancement (by up to two orders 
of magnitude) of the electrical conductivity as the iPCM undergoes the phase transition from the RESET to the SET state is the emergence of the gapless chiral monopole-antimonopole Weyl pairs. 
We further demonstrate that the shape of the Fermi surface depends sensitively on the surface termination exhibiting different topology.
The Ge-Te-terminated surface has solely Fermi arcs and the Sb-Te-terminated surface has both closed Fermi surface and open Fermi arcs. Our findings show a possibility of realizing and exploiting the Weyl state in materials without the need of particular experimental conditions such as high pressure or low temperature.

\section*{Results}
\textbf{Atomic and electronic structure of iPCM.}
Figure 1 shows the crystal structure of the (GeTe)$_2$(Sb$_2$Te$_3$)$_1$  superlattice in the highly resistive RESET state
with Sb$_2$Te$_3$-Te-Ge-Ge-Te- stacking sequence (Inverted-Petrov structure) and the highly
conductive SET state with Sb$_2$Te$_3$-Ge-Te-Ge-Te- stacking sequence (Ferroelectic-GeTe structure) along the rhombohedral [111] direction. 	
The Sb$_2$Te$_3$ block consists of one Te-Sb-Te-Sb-Te quintuplet layer.
Despite the large contrast in electrical conductivity, a sole difference in atomic structure is the stacking sequence of Ge-Te layers sandwiched by Sb$_2$Te$_3$ layers.

The MBJLDA band structures along the high-symmetry directions for the (a) Inverted-Petrov and (b) Ferro-GeTe (GeTe)$_2$(Sb$_2$Te$_3$)$_1$  superlattice are shown in Figs. 1(e) and (f), respectively.
As expected, the RESET state shows an inverted band structure at $\Gamma$ where the conduction band minimum is derived primarily from Te-$p$ states while the valence band maximum from cation (Ge or Sb) $p$ states, indicating
that the RESET state is a topological insulator in agreement with previous studies\cite{ipcmc}.
The direct energy gap of 38 meV at the $\Gamma$ point smaller by a factor of 5 than that of Sb$_3$Te$_5$\cite{Nechaev} indicates a reduction in the band inversion strength and thus proximity of the system to the boundary between topological and normal insulator phases.
Despite the conducting nature of the
surface states protected by the nontrivial band topology, the electric conduction is mostly dominated by the insulating bulk states rendering the RESET state highly resistive.
On the other hand, the emergence of ferroelectric polarization in the SET state induces a giant Rashba effect that lifts the degeneracy around $\Gamma$ and $A$. As a results, the valence band maxima and conduction band minima shift away from these symmetry points.
Despite these differences, both RESET and SET states show insulating behavior along the high symmetry lines may not explain the striking conductivity change in the phase transition. Obviously, it is natural to examine if the SET state is a WSM by searching the Weyl nodes in the entire BZ.

\textbf{Effect of Weyl semimetal phase on electrical properties.}
Interestingly, we find that the SET state is indeed a WSM with twelve gapless Weyl nodes at the Fermi level in the vicinity of the $A$-point. Figure 2(a) and (b) show three pairs of Weyl nodes located at the $k_{z} = \pi/c+\delta$ plane and the other three pairs at the $\pi/c-\delta$ plane, respectively, where $\delta=0.0143$ {\AA}$^{-1}$ is the deviation from the BZ cross section at $k_{z} = \pi/c$. The distances between the nearest and next nearest Weyl nodes are 0.036 {\AA}$^{-1}$ and 0.079 {\AA}$^{-1}$, respectively.
The two-dimensional band structure of the SET state in the $k_{x} - k_{y}$ planes at $k_{z} = \pi/c+\delta$, $k_{z}=\pi/c$, and $k_{z} = \pi/c-\delta$ are shown in Figs. 2(d), (e) and (f), respectively. One can clearly see that the SET state is gapped at the BZ boundary $k_{z}=\pi/c$, while it is gapless, having six Weyl nodes each at two planes away from the BZ boundary.
Calculated Berry curvature shows more detailed characteristics of the Weyl phase in the SET state. Each Weyl node behaves as an effective magnetic monopole in the momentum space with either positive or negative chirality, corresponding to a source or a sink of the Berry curvature, respectively. In Fig. 2(g), we show the Berry curvature with a dense $k$-point grid on the $k_y-k_z$ plane that passes through the two Weyl nodes, clearly indicating the outgoing and incoming flux of Berry curvature from one monopole (red) to the other monopole (blue).

Unlike Dirac nodes, which may easily become massive by weak perturbations, Weyl nodes are persistently massless until they are mutually annihilated. Furthermore, the chirality of Weyl nodes prohibits the back-scattering of conduction electrons from each chiral channel. The emergence of gapless nodes characterized by an exceptional high mobility, therefore, may explain the giant conductivity enhancement as iPCM undergoes the RESET to SET phase transition. By employing the maximally localized Wannier functions and the semiclassical Boltzmann transport equation in the constant relaxation time approximation, we calculate the room-temperature electric conductivity of the SET and RESET states, with the assumption that both state have the same relaxation time. Fig. 2(h) shows the ratio of the SET to the RESET state of both the in- and out-of-plane conductivity  as a function of the chemical potential. Our calculations show that for both transport directions, $\sigma_{SET} \sim 5.6 \sigma_{RESET}$ at $E=E_F$.
On the other hand, the conductivity ratio may increase by up to two orders of magnitude for p-doped cases, which commonly occur in PCM due to Ge vacancies.
Therefore, we can attribute the giant change of conductivity observed in experiment\cite{ipcmb} to the presence of Weyl nodes. To our knowledge, this is the first connection between Weyl state to the transport property of a real material.


\textbf{Fermi arcs and surface states.}
For the experimental verification of the Weyl state, we now address another key characteristic of a WSM; the presence of the Fermi arc surface states that connect the Weyl nodes in pairs in the surface BZ.
Using the Green's function method based on the tight-binding Hamiltonian generated by the Wannier projection scheme, we obtain the surface band structures of the Ge-Te-terminated (GTT) and the Sb-Te-terminated (STT) surfaces.
The band structure of the GTT surface [Fig. 3(a)] illustrates typical features of the Weyl phase, namely it is gapped along the $\bar{\Gamma}-\bar{M}$ direction while
it is gapless along the $\bar{\Gamma}-\bar{K}$ direction. This is consistent with the Fermi arcs in Fig. 3(c) connecting Weyl pairs across the $\bar{\Gamma}-\bar{K}$ axis.
On the other hand, the band structure on the STT surface [Fig. 3(b)] is not trivial as the GTT surface. The STT surface bands cross the Fermi level along both the high symmetry lines indicating a closed loop on the surface [Fig. 3(d)].
Interestingly, the STT surface indeed displays one closed loop together with six Fermi arcs which rather connect counter Weyl node pairs compared to the GTT surface. The relation between the Fermi-arc-connected pairs and the emergent closed loop is discussed below.

Due to shorter out-of-plane lattice constant (17.385 \AA) of the SET state compared to that (18.309 \AA) of the RESET state, the domain in the SET state may experience tensile uniaxial strain from surrounding domains in the RESET state.
The evolution of the surface states at the Fermi level on the GTT (STT) surface as a function of the Te-Te interlayer distance, $t$, between the Sb$_2$Te$_3$ and GeTe building blocks (Fig. 1) is shown in the top (bottom) panels Fig. 3(c) [3(d)]. The red circles denote the surface-projected Weyl nodes and the white-blue color gradient denotes spectral weight of surface states in the same scale with the band structures.
Figures 3 (c) and (d) demonstrate that the Weyl nodes of the SET state are robust under tensile strain where they mutually merge and disappear under compressive strain.
By decreasing the vacancy layer thickness $t$ by 0.1 {\AA}, the six pairs of Weyl nodes merge along the $A-H$ high symmetry directions in the bulk BZ, reopening the band gap of 7 meV.
It is worthwhile to note that, in comparison to BiTeI, the type of strain (tensile or compressive) required to shift the Weyl nodes from their creation points is opposite, namely it is compressive for BiTeI while it is tensile for Ferro.-GeTe.
The Fermi arcs appear regardless of the surface termination.

\section*{Discussion}
\textbf{Two types of Fermi arc connectivity.}
The distinct behavior of the Fermi arcs on the GTT and STT surfaces of the SET state can be understood as an adiabatic transition of surface states from the normal to the topological insulator
via the Weyl semimetal phase. For insulating nonmagnetic materials, an even (odd) number of surface bands crossing the Fermi level indicates trivial (nontrivial) topological phase.

Figure 4 shows schematically two possible transition pathways of surface states from the normal to topological insulators passing through WSM
by changing an external parameter \textit{m}. In the first type of transition [Fig. 4(a)], the initial trivial insulating phase has no surface states
along the high symmetry line connecting the two time-reversal-invariant momenta, $\lambda_1$ and $\lambda_2$, of the surface BZ. Fermi arcs subsequently emerge connecting Weyl nodes corresponding to monopole-antimonopole creation pairs; $w_1$-$w_2$ and $w_3$-$w_4$.
In the Weyl phase, where the Fermi arcs evolve as the external parameter changes, there are no surface states along the $\lambda_1$-$\lambda_2$ line until the system undergoes a transition to the topological insulator phase via the mutual monopole-antimonopole pair annihilation.
On the other hand, in the second type of transition [Fig. 4(b)], the initial trivial insulating phase exhibits two surface states crossing the Fermi level along the $\lambda_1$-$\lambda_2$ line. As the system enters in the Weyl phase, the emerging Fermi arcs connect Weyl nodes, $w_1$-$w_3$ and $w_2$-$w_4$, corresponding to
monopole-antimonopole annihilation pairs so that two Fermi surface states cross the $\lambda_1$-$\lambda_2$ symmetry direction {\it before}
the system undergoes a phase transition to the topological insulator. In this case the surface states in the Weyl phase consist of
both closed Fermi surface and open Fermi arcs. 
Note that in both transition paths, there are even (odd) number of surface state along the $\lambda_1$-$\lambda_2$ line in the normal (topological) insulator phase.
Consequently, the surface of a Weyl semimetal with Fermi arcs connecting creation (annihilation) monopole-antimonopole pairs should exhibit an even (odd) number of closed loops.
The GTT (STT) surface of the SET state of iPCM corresponds to the former (later) case. A recent study of the band-bending effect \cite{Rusinov} also showed that an emergent closed loop on the surface is accompanied by a partner exchange of Fermi arcs.

\textbf{Impilcation of photoemssion spectroscopy.} The appearance of the closed loop at the STT surface may hinder the observation of Fermi arcs in
angle-resolved photoemission spectroscopy measurements due to its close proximity in momentum space. However, even with
limited resolution, the spectral intensity on the $\bar{\Gamma}$-$\bar{M}$ lines should be twice of that on the $\bar{\Gamma}$-$\bar{K}$ lines.
It is also worthwhile to note that the distance of about 0.1 $\text{\AA}^{-1}$ of the Fermi arcs from $\bar{\Gamma}$  is larger than that of other predicted hexagonal Weyl semimetals\cite{VanderbiltPRB2014}.

\textbf{Effect of breaking time reversal on the Weyl phase.}
Weyl semimetals require breaking of either the time-reversal or inversion symmetry. We have also explored the
consequences of breaking the time reversal symmetry in the RESET phase under an external magnetic field by including
a Zeeman splitting term, $H_Z=\lambda_Z\sigma_z$ in the tight-binding Hamiltonian the effective magnetic field is along [001] direction and $\sigma_z$ is the Pauli spin operator. 
Figure 5 shows that the direct band gap of the RESET Inverted-Petrov structure decreases with increasing Zeeman splitting
and eventually closes above the critical field strength ($\lambda_Z=37.5$ meV).
The number of Weyl nodes depend on the symmetry of the system. For instance, the number of Weyl nodes in a system with $C_{3v}$ symmetry, \textit{i.e.} BiTeI or Ge$_2$Sb$_2$Te$_5$, is always a multiple of twelve.
The applied magnetic field, however, breaks the mirror symmetries and hence reduces the symmetry to $C_3$, giving rise to only six Weyl nodes.
The spin degeneracy of the initial band structure [Fig. 5(b)] is lifted due to the Zeeman term. With increasing splitting [Fig. 5(c)], a pair of valence and conduction bands eventually touch at $\Gamma$ point
and six gapless nodes persist along the $\Gamma-K$ direction at higher magnetic field.
Because of the minuscule separation of Weyl nodes in momentum space and the complicated dependence on magnetic field direction, the detailed Weyl properties with broken time-reversal symmetry are not discussed here. 
Perhaps a more practical way in introducing magnetic exchange splitting in iPCM without external magnetic field is via magnetic doping. Since there have been previous studies\cite{SongAM,TongAPL,SongJN} on magnetic doping in Ge$_2$Sb$_2$Te$_5$, it seems highly plausible to tune the exchange field by varying the magnetic impurity concentration.

\section*{Summary}
The iPCM undergoes a reversible structural transition between the RESET and the SET state.
We predict that the high conductive SET state exhibits the \textit{Weyl semimetal phase} due to the ferroelectric polarization of the GeTe block in contrast to the \textit{topological insulating phase} of the low conductive RESET state. 
The emergent twelve Weyl nodes in the SET state cause a giant enhancement in the conductivity by a factor of 5 $\sim$ 100 in comparison to the RESET state.
To the best of our knowledge, this is the first memory device in which the conductivity contrast between the SET and the RESET state is attributed to the emergent Weyl nodes.
Our results, implying a robust Weyl phase in the SET state, open a way to functionalize the iPCM and design novel devices. We also predict two types of Fermi arc connectivity in the SET state depending on the surface termination.

\section*{Methods}
\textbf{Calculation details.}
The density functional theory calculations were carried out using the Vienna \emph{ab initio} simulation package (VASP) \cite{Kresse96a,Kresse96b}. The pseudopotential and wave functions are treated within the projector-augmented wave (PAW) method \cite{Blochl94,KressePAW}. Structural relaxations were carried using the generalized gradient approximation as parameterized by Perdew {\it{et al.}} \cite{PBE} with van der Waals corrections\cite{Grimme}. The plane wave cutoff energy is 400 eV and a 15 $\times$ 15 $\times$ 5 $k$ point mesh is used in the BZ sampling.
The electronic properties such as the band structures, surface states, and conductivities have been calculated using the tight-binding parameters obtained from VASP-WANNIER90 calculations \cite{Mostofi}. In the Wannier function projection, the Modified Becke-Johnson Local Density Approximation (MBJLDA) \cite{Tran_mBJ,Koller_mBJ} potential was employed, which has been shown to yield accurate band gaps, effective masses, and frontier-band ordering. Spin-orbit coupling was included in the projection with a 10 $\times$ 10 $\times$ 4 $k$-point mesh in the BZ sampling.
The Weyl nodes were traced using a steepest decent method while the longitudinal electric conductivity was calculated within the semiclassical Boltzmann transport theory\cite{Mostofi}.

\begin{acknowledgments}
The work at CSUN is supported by NSF-Partnership in Research and Education in Materials (PREM) Grant DMR-1205734 and by the US Army of Defense (DOD), Grant Contract number W911NF-16-1-0487.
Y.-S. S. and S.-H. J. were supported by the National Research Foundation of Korea through SRC program (Contract No. 2011-0030046).
J. Kim and R. Wu were supported by DOE-BES (Grants Nos. DE-FG02-05ER46237 and SC0012670). Calculations were performed on parallel computers at NERSC supercomputer centers.
\end{acknowledgments}

\section*{Author contributions}
J.K. and N.K. designed this project. J.K., J.K., and Y.-S.S. performed the calculation. R.W., S.-H.J., and N.K. wrote the manuscript. All authors commented on the manuscript and the results.

\begin{figure*} [p]
	\includegraphics[width=15cm]{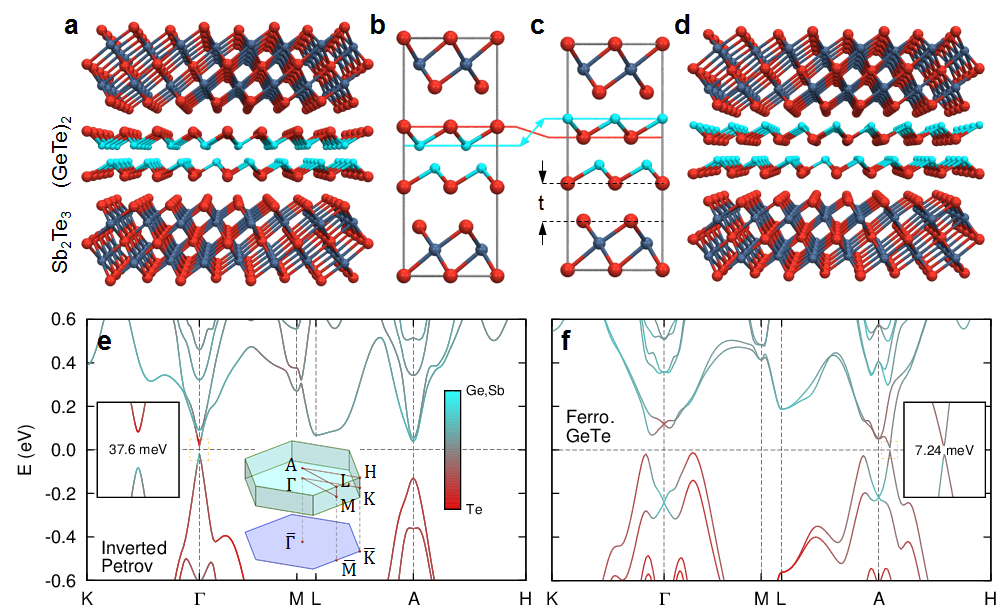}
	\caption{\textbf{Atomic and electronic structures of iPCM in the RESET and SET states.} Crystal structure of the (GeTe)$_2$(Sb$_2$Te$_3$)$_1$ superlattice in the (\textbf{a},\textbf{b}) RESET Inverted-Petrov structure state (Sb$_2$Te$_3$-Te-Ge-Ge-Te- stacking)  and (\textbf{c},\textbf{d}) SET ferroelectic structure (Sb$_2$Te$_3$-Ge-Te-Ge-Te- stacking).
	MBJLDA bulk band structure along symmetry lines of (\textbf{e}) Inverted-Petrov and (\textbf{f}) Ferro-GeTe crystal structures with the Fermi level set at 0 eV.
	The bands denoted with red (cyan) derive from anion (cation) states.
	Inset shows the bulk BZ with the high symmetry points and the projected 2D BZ for the (001) surface.}
\end{figure*}

\begin{figure*} [p]
	\centering
	\includegraphics[width=15.0cm]{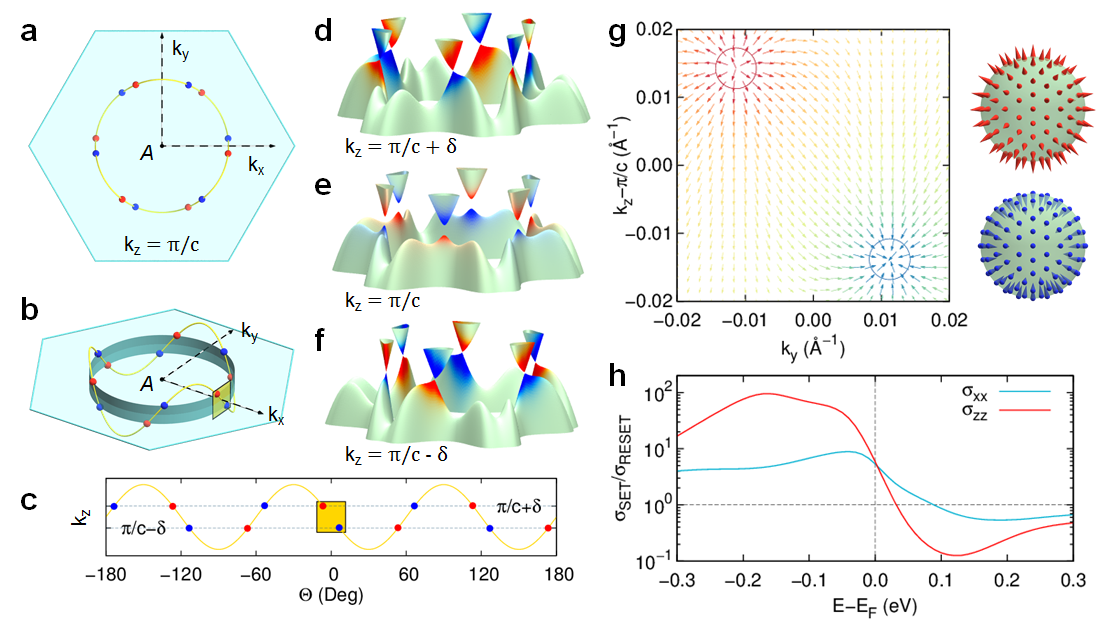}
	\caption{\textbf{Weyl node assisted conductivity enhancement in the SET state.} (\textbf{a}) Top view and (\textbf{b}) 3D view of locations of Weyl nodes in the BZ of the Ferro-GeTe structure, where red (blue) circles denote Weyl nodes with positive (negative) chirality. The hexagon denotes a cross section of BZ at $k_{z}=\pi/c$. The position of Weyl nodes relative to $A$ point is scaled five-fold bigger for visibility. (\textbf{c}) Location of Weyl nodes along the k$_z$ direction versus the azimuthal angle $\theta$ in the k$_x$-k$_y$ plane. Yellow line is a fitted sinusoidal curve.
	(\textbf{d}-\textbf{f}) Calculated bulk band structures of Ferro-GeTe structure on the 2D $k_{x}$-$k_{y}$ plane at (\textbf{d}) $k_{z} = \pi/c+\delta$, (\textbf{e}) $\pi/c$, and (\textbf{f}) $\pi/c-\delta$, respectively, where $\delta=0.0143$ $\AA^{-1}$ is the deviation of Weyl nodes from the BZ cross section. Red and blue colors in \textbf{e} denote the $\hat{z}$ component of Berry curvature, $\Omega_{z}$, while those in \textbf{d} and \textbf{f} denote the chirality, $\sum_{i}{\Delta\mathbf{k}_{i}\cdot\mathbf{\Omega}(\mathbf{k})/(|\Delta\mathbf{k}_{i}|\text{exp}(|\Delta\mathbf{k}_{i}|/r_{0})}$. Here $\Delta\mathbf{k}_{i}=\mathbf{k}-\mathbf{k}_{i}$, where $\mathbf{k}_{i}$ is each Weyl node position and $r_0$ is an arbitrary cutoff radius. (\textbf{g}) Distribution of Berry curvature on a part of $k_{y}$-$k_{z}$ plane (the square in \textbf{b} highlighted in yellow).
	3D view of hedgehog-like Berry curvature near two selected Weyl points.
	(\textbf{h}) Calculated conductivity ratio of the SET to RESET states as a function of the chemical potential. }
\end{figure*}

\begin{figure*} [p]
	\includegraphics[width=18cm]{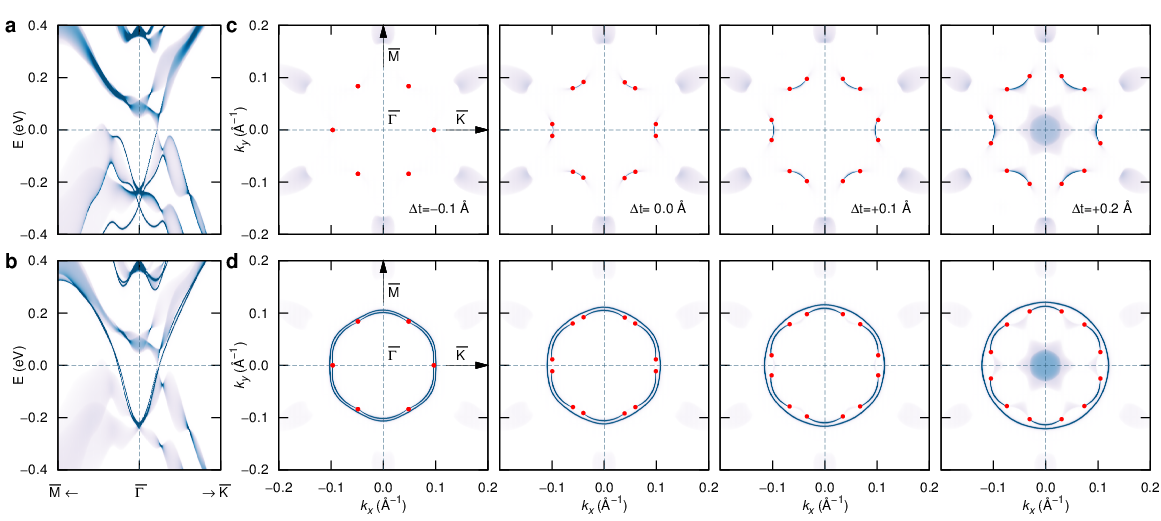}
	\caption{\textbf{Surface states of the semi-infinite iPCM structure in the SET state.} Surface band structures for (\textbf{a}) GTT surface and
	(\textbf{b}) STT surface with $\Delta t = t- t_{eq}$=0, where $t_{eq}$ is the equilibrium Te-Te interlayer distance between the Sb$_2$Te$_3$ and GeTe building blocks.
	(\textbf{c}) Fermi surfaces on the (001) GTT surface for different values of $t$. (\textbf{d}) Fermi surfaces on the (001)
	STT surface for different values of $t$.
	Red dots in \textbf{c} and \textbf{d} denote Weyl nodes projected on the surface BZ.}
\end{figure*}

\begin{figure} [p]
	\includegraphics[width=8.4cm]{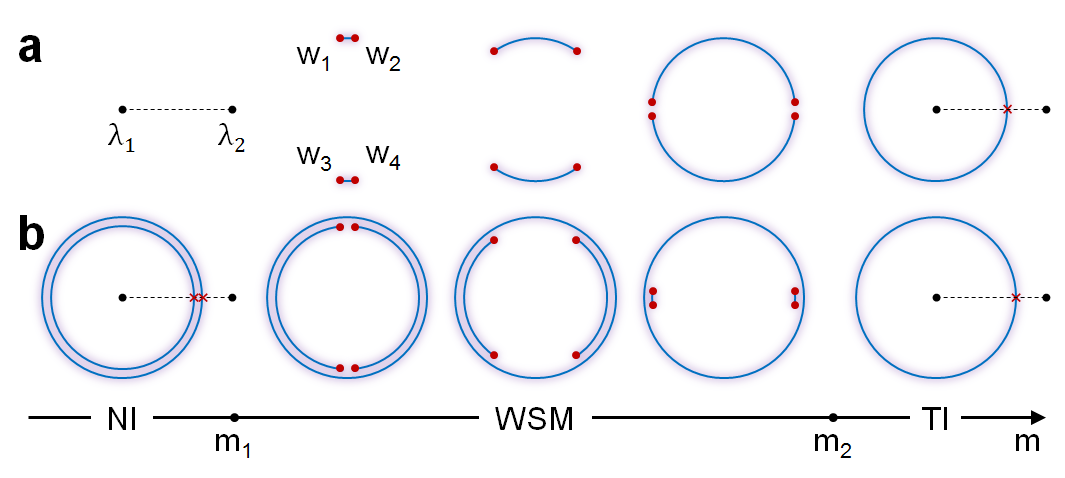}
	\caption{\textbf{Schematic view of Fermi arc connectivity.} Two types of surface-state-evolution as the system undergoes a topological phase transition from normal insulator to Weyl semimetal to topological insulator phases. (\textbf{a}) Fermi arc connecting creation pair. (\textbf{b}) Fermi arcs connecting annihilation pair. $\lambda_i$ denotes time-reversal-invariant momenta on the surface BZ and $w_i$ are the Weyl nodes.}
\end{figure}

\begin{figure} [t]
	\includegraphics[width=7.6cm]{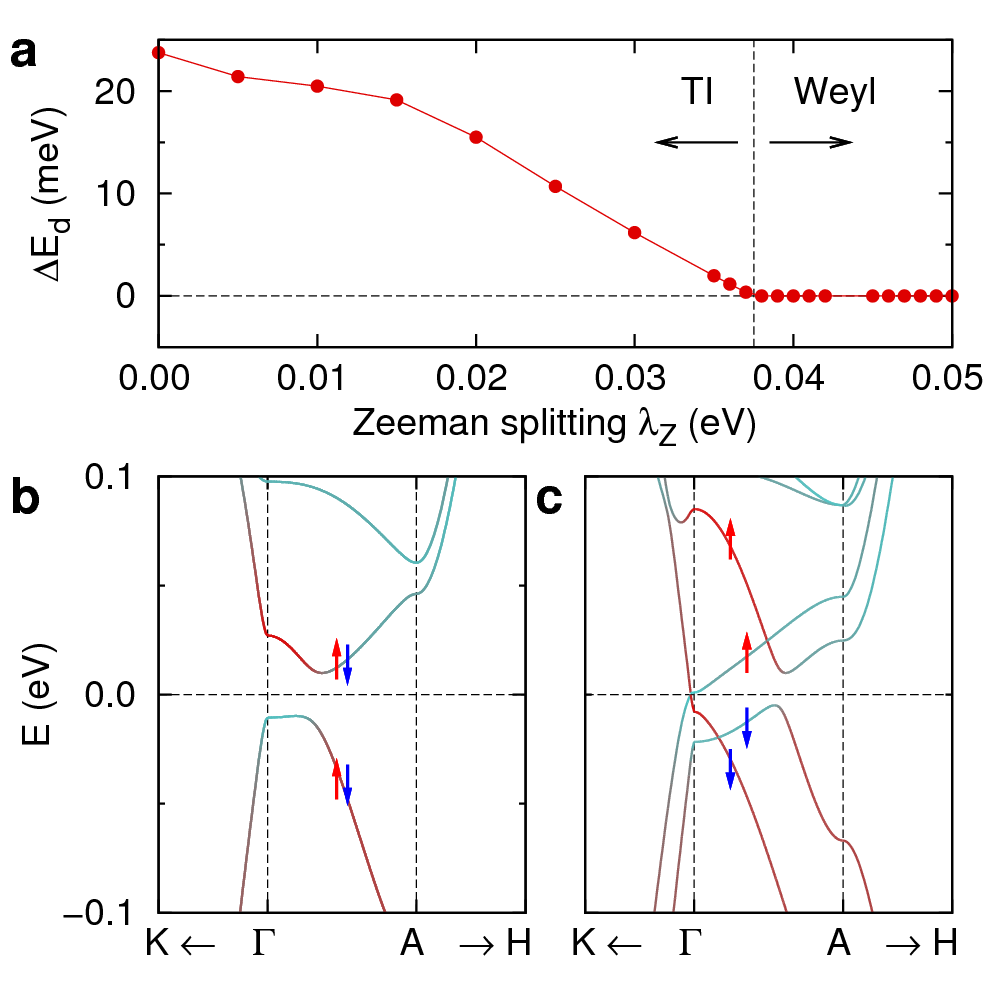}
	\caption{\textbf{Magnetic-field-induced emergence of Weyl phase in the RESET Inverted Petrov structure.} (\textbf{a}) Direct band gap as a function of Zeeman splitting, $\lambda_Z \sigma_{z}$, where $\sigma_{z}$ is the Pauli spin operator. Calculated band structures with (\textbf{b}) $\lambda_Z=0$ and (\textbf{c}) $\lambda_Z=0.05$ eV. Color of the bands denotes the orbital character as in Fig. 2.}
\end{figure}

\end{document}